\documentstyle[aps,pre]{revtex}               
\tighten

\begin{document}
\draft
\twocolumn

\title{Damage spreading during domain growth}
\author{I. Graham$^{1,2,3}$, E. Hern\'andez-Garc\'\i a$^3$, and
Martin Grant$^2$}

\address{$^1$ Industrial Materials Institute, National Research
Council of Canada\\
75 De Mortagne Blvd, Boucherville, Qu\'ebec, Canada, J4B 6Y4 }

\address{$^2$ Center for the Physics of Materials and Department of
Physics\\
McGill University, 3600 University Street, Montreal, Qu\'ebec,
Canada, H3A 2T8 }

\address{$^3$ Departament de F\'\i sica, Universitat de les Illes
Balears\\
E-07071 Palma de Mallorca, Spain}

\date{March 23, 1994}

\maketitle
\narrowtext

\begin{abstract}
We study damage spreading in models of two-dimensional systems
undergoing first order phase transitions.  We consider
several models from the same non-conserved order
parameter universality class, and find unexpected
differences between them. An exact solution of the
Ohta-Jasnow-Kawasaki model yields the damage growth law
 $D \sim t^{\phi}$, where $\phi = t^{d/4}$ in $d$ dimensions. In contrast,
 time-dependent Ginzburg-Landau simulations and Ising simulations
 in $d= 2$
 using heat-bath dynamics show power-law growth, but with an
 exponent of approximately $0.36$, independent of the system
 sizes studied.
 In marked contrast, Metropolis dynamics shows damage growing via
 $\phi \sim 1$, although the damage
 difference grows as $t^{0.4}$.
\end{abstract}

\pacs{64.60.-i, 05.50.+q}


The time dependence of fluctuations and correlations is a central
aspect of the study of equilibrium and non-equilibrium phenomena.
Fluctuations in the time-evolution of a dynamical system have two
different origins, arising either from the random assignment of the
initial conditions or from random influences during the dynamics --
i.e. the (thermal) noise. A particularly interesting method for
isolating the influence of initial conditions on the subsequent
evolution is {\it damage spreading}\cite{kauf69}.
Consider a kinetic Ising model with site spins $\pm1$,
evolving via Monte Carlo dynamics.  We introduce damage as
follows.  Starting with two lattices with identical spin
configurations we take one system, select at random a single
site and flip the spin.  This introduces a
microscopic fluctuation into the `damaged' system relative
to the undamaged one.  Both lattices, including the damaged
site, are then evolved according to the same Monte Carlo
dynamics, by using the same sequence of random numbers for
both.  Any subsequent difference between the two is due
to the difference in the initial conditions, since without
the initial damage the two systems are always identical.

One common measure of damage between two Ising systems
is the Hamming distance between the configurations, given by
\begin{equation}
\label{damage}
D(L,t) \equiv \frac {1} {2} \sum_i | S_i(t) - S_i^D(t) |
\end{equation}
where $S_i$ is the state of spin $i$ in the undamaged lattice,
$S_i^D$ is the state of site $i$ in the damaged lattice
and $N=L^d$ is the number of spins in a hypercubic lattice of size
$L$.  Another important measure is the damage difference,
defined by
\begin{equation}
\label{damdiff}
\Gamma(L,t) \equiv \frac {1} {4} \sum_i \left( S_o(0) - S_o^D(0) \right)
                                        \left( S_i(t) - S_i^D(t) \right).
\end{equation}
This measures the difference between the amount of damage that is
the same as that initially imposed at site $o$ and time $t=0$ and
the amount that is of opposite orientation. The term
$\left( S_o(0) - S_o^D(0)\right)$ defines the `parity' of the initial
damage at site $o$ and time $t=0$. We say that damage is positive
if it is the same as that initially imposed and negative otherwise.
At equilibrium $\Gamma(t)$ can be related to
thermodynamic quantities such as the susceptibility and correlation
functions, provided the damaged site is frozen for all time in its
$t=0$ configuration \cite{coni89,herr90}.
There are no known relationships between damage and correlations
when the damaged site is allowed to evolve, which is the method
considered here.  Lastly we remark that damage evolution in systems far
from equilibrium has been little studied, and that the relationship
between damage growth and traditional dynamical properties in these
cases is unknown.

In this paper we examine this latter situation, namely the evolution
of damage in non-equilibrium systems, specifically during domain growth.
 Following a quench from a disordered state to a low-temperature state
 where phases coexist, a system dynamically evolves by the formation and
 subsequent growth of domains of ordered phase.  The average size of
 those domains grows as $R \sim t^n$, for late times where $n$ is the
 growth exponent, and the morphological evolution involves scale
 invariance, where all time dependence enters through the characteristic
 diverging length $R$.  An important issue in this area is universality:
 what are the common features which characterize a universality class,
 like $n$, and how are universality classes determined?  The most
 well established universality class in domain growth
 is model A \cite{hohenberg},
 where the scalar order parameter is nonconserved, and \cite{gunton}
 $n = 1/2$.
 Systems in this class include binary alloys undergoing order-disorder
 transitions, as well as the ferromagnetic spin-flip Ising model, and
 continuum models such as the time-dependent Ginzburg-Landau equation
 (TDGL) \cite{gunton} and the Ohta, Jasnow and Kawasaki (OJK) model
 \cite{ojk}.  Herein, we investigate damage evolution during domain
 growth in these latter three models.  We find some similarities, but
 important differences, for the different models.
 Hence we  suggest that universality in domain growth does not include
 damage evolution.

We first generalize the concept of damage to allow for non-discrete
variables. With this definition we are able to show
 that OJK dynamics leads to $D \sim t^{\phi}$, with a damage
 exponent $\phi = d/4$ in $d$ dimensions.
 Two-dimensional simulation studies of the TDGL equations and of
 the Ising model using heat-bath dynamics both yield power-law growth for
 damage.  However, $\phi \approx 0.36$, significantly smaller than the
 OJK model. More remarkably, Ising simulations using Metropolis dynamics
 in $d = 2$ show  $\phi \sim 1$.  It should be noted that
in all the models studied damage disappears with high probability at
long times. But for some realizations of the initial conditions and
thermal noises, it attains macroscopic size. The power laws above
apply to the damage averaged over samples of the full ensemble of
realizations.


The OJK model\cite{ojk} was introduced as an approximation
to the Allen-Cahn equation describing interface motion during phase
ordering with a non-conserved order parameter. It is considered
an important model in its own right \cite{young} as it readily yields
analytic predictions for quantities such as the growth exponent and
scaling functions.

Consider the evolution of a system described by an order
parameter $\psi({\bf x},t)$ which takes the value $\psi_{eq}$ in one
of the phases and $-\psi_{eq}$ in the other. In the OJK model
the evolution of $\psi({\bf x},t)$ is given by
\begin{equation}  \label{sign}
\psi({\bf x},t)=\psi_{eq}{\rm sign}(u({\bf x},t))=
    \psi_{eq}\frac {u({\bf x},t)} {| u({\bf x},t) |}\ ,
\end{equation}
where the auxiliary field $u({\bf x},t)$ satisfies the diffusion equation:
$\dot u({\bf x},t) = \nabla^2 u({\bf x},t)$.
Without loss of generality the values of $\psi_{eq}$ and of the diffusion
coefficient
are taken here to be unity.
The initial condition
$u_0({\bf x}) \equiv u({\bf x},t=0)$
is
usually taken as a random Gaussian field of zero average and short-range
correlations.
The exact solution of the diffusion equation
is
$u({\bf x},t) = \int d{\bf x} G({\bf x} - {\bf x} ',t) u_0({\bf x} ')$,
where  G({\bf x} ,t) is a Green function.
Substitution of
this expression into Eq.\ (\ref{sign}) leads to
an equation of evolution for $\psi({\bf x},t)$  representing the growth of
domains of typical size $R(t) \sim t^{1/2}$ \cite{ojk}.

We need continuum definitions for the damage measure $D(L,t)$
and for the initial damage consistent with those used
for the Ising model. We define the initial damage in terms of the
initial unperturbed
field $u_0({\bf x})$. Several definitions of the `damaged'
field $u_0^D$ are possible: we found all to give essentially
identical results. We present detailed results for the case in which
$u_0^D({\bf x}) \equiv - u_0({\bf x})$ in a damaged region of volume $V_D$ and
$u_0^D({\bf x}) \equiv  u_0({\bf x})$ outside. We define the damage measure by
\begin{eqnarray}
\label{damcontinuous}
D(L,t) &\equiv& \frac {1} {2} \int_{L^d} d {\bf x} \ |\psi({\bf x},t)
       - \psi^D({\bf x},t)| \\
&=&
\frac{1}{2} \left( L^d -
  \int_{L^d} d {\bf x} \  \psi({\bf x},t) \psi^D({\bf x},t) \right).
\end{eqnarray}
The calculation of the average damage (ensemble-averaged over different
realizations of the random initial conditions) then reduces to the
calculation of a correlation function and a subsequent integration.
The detailed calculation will be presented elsewhere: here we quote
the main results.

In the regime of long times
and large $L$ and with the initially damaged volume much smaller
than system size we find the following scaling result
\begin{equation}
\label{damscaling}
\left< D(L,t) \right> = L^{\frac{d}{2}}  F(t/L^2),
\end{equation}
where the scaling function $F$ is
\begin{equation}
\label{Function}
F(\tau) = \frac{ 2 \sqrt{V_D} } {\pi}
\frac {1} {\sqrt{G({\bf 0},2\tau)}}
\int_{I^d} d{\bf x} \ G({\bf x},\tau) .
\end{equation}
$I^d$ is the the unit hypercube.
  The crucial and disputable element here is the $L^{d/2}$ factor.
  This arises from the assumption of Gaussian de-coupling in the OJK model.
The scaling form in the limit $t\ll L^2$ is
\begin{equation}
\label{limit1} \left< D(t \ll L^2) \right>  \sim t^{d \over 4}.
\end{equation}
Plots of $\left< D(L,t) \right>$ from Eqs.\ (\ref{damscaling}) and
(\ref{Function}) for
$d=2$ are shown in scaled form in Fig.\ \ref{fig:ojk}.
The power-law regime is clearly seen, as are the corrections
to scaling at short times.

We also numerically iterated the OJK equations on a lattice to visualize
the evolution of the damage.  We found that damage, when
it exists, evolves along the interface between the two
phases. Damage then either disappears as the interfaces disappear,
or it grows to encompass the entire system. In addition we confirmed
that only one parity of damage dominates: in particular if we damage
only a single lattice site the propagated damage is always
of the same parity as initially imposed. This result can be verified
analytically.  Therefore in OJK, with this
damage prescription, the damage difference and Hamming distance are
identical.


The OJK damage definition (Eq.\ (\ref{damcontinuous})) also applies
to the TDGL case. The initial damaged configuration in this case was
chosen to be $\psi^D({\bf x},t=0)=-\psi({\bf x},t=0)$ inside a small
damaged
region and $\psi^D({\bf x},t)=\psi({\bf x},t=0)$ outside, where again
$\psi({\bf x},t=0)$ is a Gaussian random field with short
range correlations.
Both $\psi$ and $\psi^D$ are evolved according to the TDGL equation:
\begin{equation}
\label{langevin}
\dot \psi({\bf x},t) = \psi({\bf x},t) + \nabla^2 \psi({\bf x},t) -
\psi({\bf x},t)^3.
\end{equation}
Addition of a noise term does not significatively change the results.
The average damage evolution is shown in
Fig.\ \ref{fig:heat_bath_1}. The power-law damage growth regime is clearly
present and persists up until the system magnetization begins to
saturate.  We obtain $\phi = 0.36 \pm 0.02$,  which is
 smaller than the OJK prediction of $d/4= 1/2$.  The errors are estimated
 by one standard deviation of the statistics of the data.

As with the OJK model, we found that the TDGL equations,
solved on a discrete lattice, only propagate damage of the
same parity as that initially imposed, provided we only damage a single
lattice site. If we damage a larger domain there is transient regime
during which both parities of damage compete, but eventually one
disappears and damage follows $\phi \approx 0.36$.


Ising simulations, using both heat-bath and Metropolis dynamics,
were performed on two-dimensional square lattices, using a multi-spin
coding algorithm described previously \cite{multi_spin,ian92}.
However the block-spin coding trick \cite{multi_spin} was avoided, as
some kinds of parallel updating have been found to affect damage
dynamics \cite{nobre92,serial}.  Spins to be evolved were
chosen at random from amongst all spins, a single time step
(mcs/s) being $L^2$ such attempts. The initial state was an infinite
temperature equilibrium configuration.

Typical results for damage evolution via heat-bath dynamics are
shown in Fig.\ \ref{fig:heat_bath_1}, for two system sizes and
for temperatures $T = 0.8 T_c$ and $T = 0$.   The early-time
transient is very short, and is quickly followed by a slow
power-law growth regime.
The power-law exponents are $0.36\pm0.02$, identical with the
TDGL results and smaller than the OJK prediction of $1/2$.
It also appears that the system size does not strongly affect this
exponent: systems of size $L=48$ have essentially the same slope
as those four times as big. There is no evidence for faster
growth at late times: the growth simply slows as finite-size
effects are encountered, and then saturates.
Temperature also does not appear to affect the exponent in the
power-law regime. In
general increasing the temperature decreases the total
damage. This is consistent with previous studies of damage
in equilibrium systems using the heat-bath algorithm.

 We cannot rule out the possibility that the $0.36$ exponent is only
 effective, and that systematic errors preclude our observation of
 $d/4$.  Our estimated errors are statistical, and comparable
 systematic errors are possible.  However, we have made a careful study
 of systems of different sizes, and our numerical results for $\phi$
 are not consistent with the exact result of the OJK model of $d/4$.

It was previously demonstrated that the heat-bath algorithm
can only propagate damage of the same parity as that initially
imposed\cite{coni89}.  Therefore as in the previous two cases
the Hamming measure and the damage difference are identical.

We obtain remarkably different behavior using the
Metropolis algorithm. Typical Metropolis dynamics results are
shown in Fig.\ \ref{fig:metropolis1}, where again we have plotted
results demonstrating the effect of varying system size
and temperature.  The results are different
 from all results presented above.
Even at earliest
times the growth of damage is {\it faster} than $t^{1/2}$, with this
exponent increasing with time.  At intermediate times this
damage saturates at $\phi \sim 1$, i.e. approximately
twice as fast as seen by the
other dynamics. Simulations to much later times indicate no
deviation from this dependence until
finite-size effects are encountered and growth stops.

Equivalence between the three numerical models can be recovered if we
consider the damage difference, Eq.\ (\ref{damdiff}).  A typical
plot of the damage difference is shown in Fig.\ \ref{fig:metropolis1}
for systems of size $L=48$.
Although quite noisy, these data clearly show a power-law growth
slower than $t^{1/2}$.  A best fit yields $0.40\pm 0.04$, consistent
with the TDGL and Ising heat-bath results. Of course we again cannot rule
out that the possibility that this exponent approaches 0.5 for
larger systems, indicating strong finite-size effects.  However,
we see no evidence for such systematic errors in the range of sizes
studied ($L=48$ to $L=192$).

The reason for the surprising differences between the Ising simulations
lies, at the
microscopic level, in  the difference between the two dynamics.
Metropolis dynamics allows both parities of damage to exist.
By visualizing the damage evolution we find that Metropolis
dynamics produces significant amounts of both damage.
Individually averaging these two parities of damage shows that
both are dominated by $\phi \sim 1$.  Damage of the {\it
same} parity as the initial damage shows a large initial
transient related to the early-time growth
of the damage difference, before crossing over to the
faster growth, while the opposite-parity damage quickly assumes
$\phi \sim 1$.

In conclusion, we have
 presented numerical evidence
that non-equilibrium damage spreading
lacks the universality of domain growth. In particular we have shown
that the TDGL model and the Ising models are equivalent, but only
when the damage difference is used as the measure.   More importantly
our work suggests that the damage algorithm breaks the universality
between the TDGL, Ising, and OJK models.
More detailed comparisons between these models, including
the consideration of finite-size effects, finite-size scaling, and the
full derivation of the OJK analytical results will be presented in a
future publication.

E.H-G acknowledges support from Direcci\'on General de Investigaci\'on
Cient\'\i fica y T\'ecnica (Spain), Project No.  PB92-0046-C02-02.


\begin{figure}
\caption{The $d=2$ OJK prediction
for a initially damaged region of $V_D=4$ and three
different system sizes $L$. The scaling region and the
damage growth via
$t^{d/4}$ are evident, as are corrections
to scaling at short times.}
\label{fig:ojk}
\end{figure}

\begin{figure}
\caption{Damage evolution for a $2d$ quenched Ising
model using heat-bath dynamics, showing the effect of temperature
and of system size.  The growth of damage is evidently
algebraic, even for the smaller systems ($L=48$). The exponents are
$0.36 \pm 0.01$ for $L=48$, $T=0.8T_c$ (24576 runs averaged);
$0.37 \pm 0.03$ for $L=192$, $T=0.0T_c$ (3072 runs averaged) and
$0.36 \pm 0.01$ for $L=192$, $T=0.8T_c$ (3072 runs averaged).
The dotted line shows the TDGL damage evolution averaged over
6400 realizations of the initial conditions, arbitrarily scaled to
fit on the same scale as the Ising data. The system size is
$L=90.5$ and the initially damaged volume is $V_D=0.5$. Note the
discrepancy with the $t^{d/4}=t^{1/2}$ OJK prediction. }
\label{fig:heat_bath_1}
\end{figure}

\begin{figure}
\caption{Damage evolution for a $2d$ quenched Ising
model using Metropolis dynamics, showing the effect of temperature
and of system size.  The regimes for damage
are only evident
for the larger system, and at late times.   The solid line shows
a damage evolution $\propto t$.  The damage
difference is also plotted, and the slow-growth behavior is
evident.  The late-time tangent to the growth gives an exponent
of $\phi  = 0.40 \pm 0.04$.}
\label{fig:metropolis1}
\end{figure}


\begin{thebibliography}{99}

\bibitem{kauf69} S.A.\ Kauffman, J.\ Theor.\ Biol.\ {\bf22}, 437 (1969);
B.\ Derrida and Y.\ Pomeau, Europhys.\ Lett.\ {\bf 1}, 45 (1986);
B.\ Derrida, Philos.\ Mag.\ B {\bf 56}, 917 (1987); D.\ Stauffer,
Philos.\ Mag.\ B {\bf 56}, 901 (1987);
B.\ Derrida and G.\ Weisbuch, Europhys.\ Lett.\ {\bf 4},
657 (1987);
H.E.\ Stanley, D.\ Stauffer, J.\ Kert\'esz and H.J.\
Herrmann, Phys.\ Rev.\ Lett.\ {\bf 59}, 2326 (1987).

\bibitem{coni89} A.\ Coniglio, L.\ de Arcangelis, H.J.\ Herrmann and
N.\ Jan, Europhys.\ Lett.\ {\bf 8}, 315 (1989).

\bibitem{herr90} H.J.\ Herrmann in {\it Computer Simulation Studies in
Condensed Matter Physics II: Springer Proceedings in Physics, Vol.\ 45},
edited by D.P.\ Landau, K.K.\ Mon and H.-B,\ Sch\"uttler (Springer-Verlag,
Berlin, 1990);
K.\ MacIsaac and N.\ Jan (unpublished);
S.C.\ Glotzer, P.H.\ Poole and N.\ Jan, J.\ Stat.\ Phys.
{\bf 68}, 895 (1992);
G.\ Batrouni and A.\ Hansen, J. Phys. A {\bf 25}, L1059
(1992).

\bibitem{hohenberg} P.C. Hohenberg and B.I. Halperin, Rev. Mod. Phys. {\bf
49}, 435 (1977).

\bibitem{gunton} J.D. Gunton, M. San Miguel and P.S. Sahni, in {\sl Phase
Transitions and Critical Phenomena, Vol.8}, edited by C. Domb and J.L.
Lebowitz (Academic, London, 1983).

\bibitem{ojk} T. Ohta, D. Jasnow, and K. Kawasaki, Phys. Rev. Lett. {\bf 17},
1223 (1982).

\bibitem{young} C. Yeung and D. Jasnow, Phys. Rev. B {\bf 42}, 10523 (1990).

\bibitem{multi_spin} G.\ Bhanot, D.\ Duke and R.\ Salvador,
J.\ Stat.\ Phys.\ {\bf 44}, 985 (1986); C.\ Roland and
M.\ Grant, Phys. Rev.\ B. {\bf 39}, 11971 (1989).

\bibitem{ian92} I.S.\ Graham and M.\ Grant, J.\ Phys.\ A, L1195 (1992).

\bibitem{nobre92} E.T. Gawlinski, M. Grant, J.D. Gunton, and K. Kaski, Phys.
Rev. A {\bf 31}, 281 (1985);
F.D.\ Nobre, A.M.\ Mariz and E.S.\ Sousa, Phys.\ Rev.\ Lett.\
{\bf 69}, 13 (1992).

\bibitem{serial} We found that the vectorized (block-spin coding) and
non-vectorized codes gave slightly different asymptotic damage values,
in keeping with the notion that this quantity is nonuniversal and can
depend on the implementation of the dynamics. Other quantities, such
as the scaling exponents, were unaffected.

\end{thebibliography}
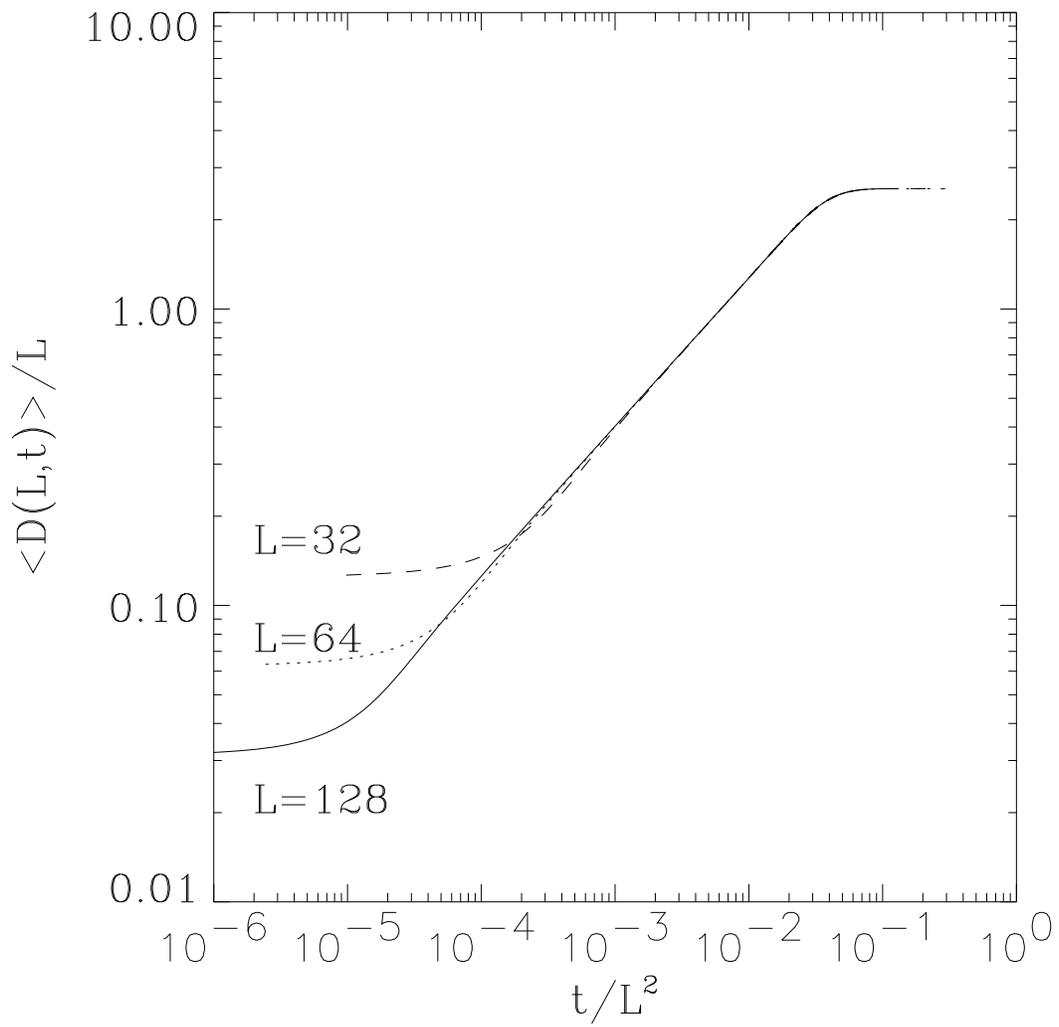
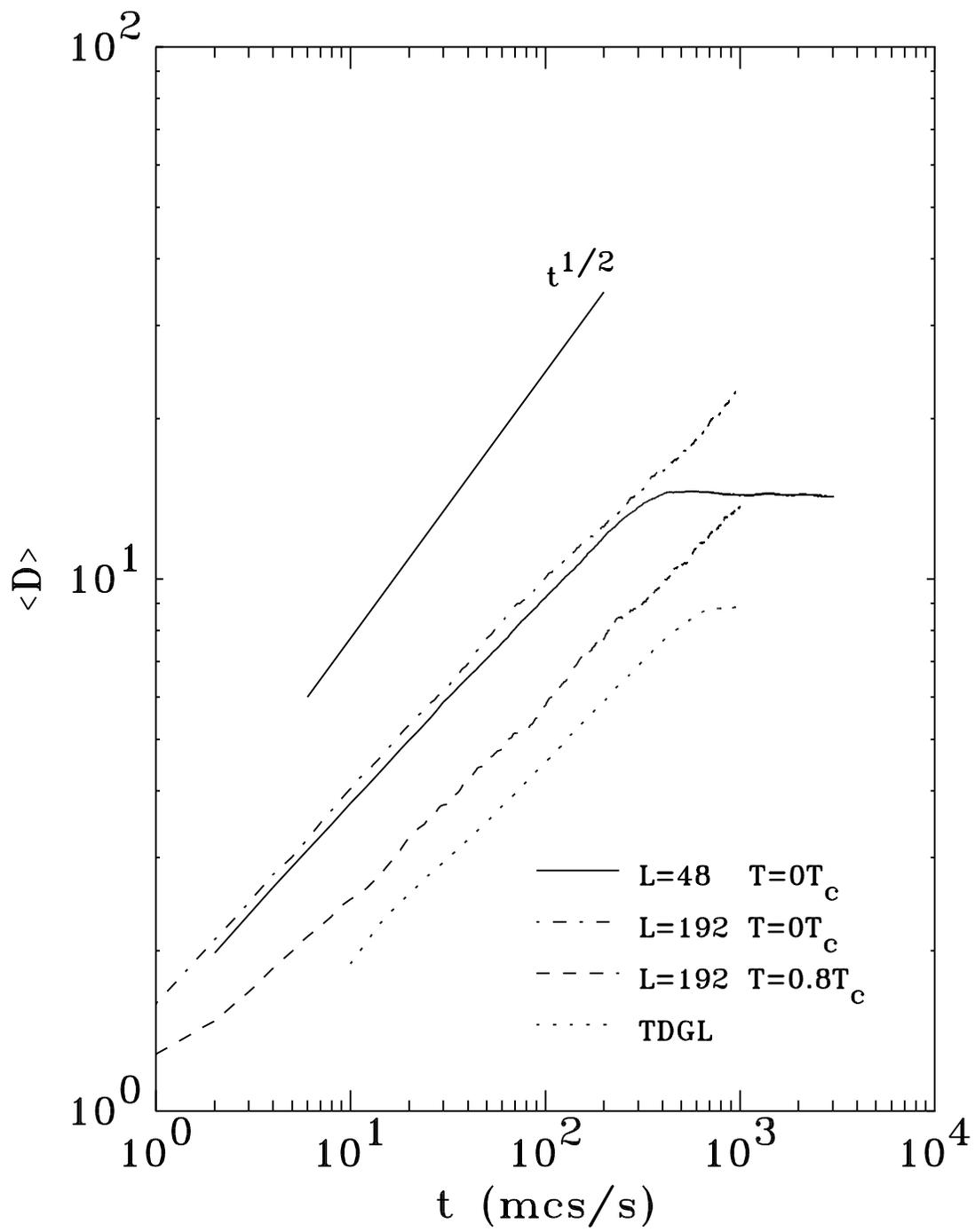
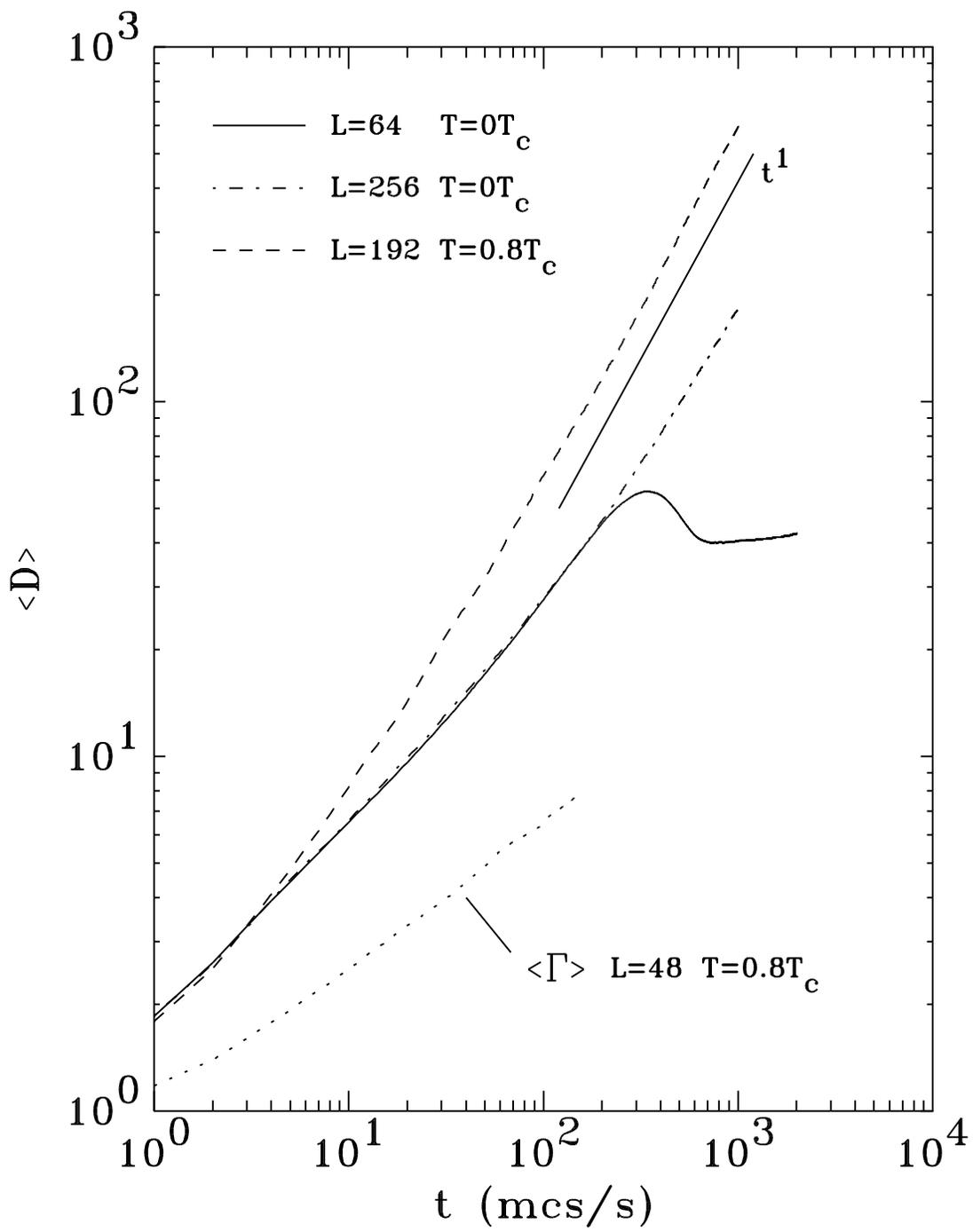
\end{document}